\documentclass[conference]{IEEEtran}

\usepackage{array}
\usepackage{makecell}
\usepackage{graphicx}

\usepackage{amsmath,amssymb,amsfonts}
\usepackage{algorithmic}
\usepackage{graphicx}
\usepackage{textcomp}
\usepackage{xcolor}
\usepackage{mathtools}
\usepackage{hyperref}

\usepackage{multirow}
\usepackage{makecell}

\usepackage{tikz}
\usepackage{tabularx}
\usepackage{svg}
\usepackage{float}

\usepackage{cleveref}
\usepackage{tcolorbox}
\usepackage{dialogue}

\usepackage{scrextend}

\AtBeginDocument{%
  \providecommand\BibTeX{{%
    \normalfont B\kern-0.5em{\scshape i\kern-0.25em b}\kern-0.8em\TeX}}}

\newcolumntype{M}[1]{>{\centering\arraybackslash}m{#1}}
\newcolumntype{N}{@{}m{0pt}@{}}

\newfloat{excerpt}{h}{lob}
\floatname{excerpt}{Excerpt}
\Crefname{excerpt}{Excerpt}{Excerpts}
\newcolumntype{Y}{>{\centering\arraybackslash}X}

\begin{document}

\title{Dialogue Management for Interactive API Search}

\author{\IEEEauthorblockN{Zachary Eberhart and Collin McMillan}
\IEEEauthorblockA{\textit{Department of Computer Science and Engineering} \\
\textit{University of Notre Dame}\\
Notre Dame, USA \\
\{zeberhar, cmc\}@nd.edu}
}


\maketitle

\begin{abstract}
API search involves finding components in an API that are relevant to a programming task.  For example, a programmer may need a function in a C library that opens a new network connection, then another function that sends data across that connection. Unfortunately, programmers often have trouble finding the API components that they need. A strong scientific consensus is emerging towards developing interactive tool support that responds to conversational feedback, emulating the experience of asking a fellow human programmer for help.  A major barrier to creating these interactive tools is implementing dialogue management for API search. Dialogue management involves determining how a system should respond to user input, such as whether to ask a clarification question or to display potential results.  In this paper, we present a dialogue manager for interactive API search that considers search results and dialogue history to select efficient actions. We implement two dialogue policies: a hand-crafted policy and a policy optimized via reinforcement learning. We perform a synthetics evaluation and a human evaluation comparing the policies to a generic single-turn, top-N policy used by source code search engines.



\end{abstract}


\section{Introduction}

Application Programming Interface (API) search involves finding components of an API that are relevant to a programming task.  Programmers often find themselves in a situation in which they know that an API implements a particular functionality, but they do not know exactly which functions of the API to use to accomplish that functionality.  For example, a programmer may know that {\small \texttt{libssh}}~\cite{libsshdocs} creates and controls SSH network connections, but he or she may not know exactly which functions to use to actually open a new connection and send data to the remote host. This problem is the motivation behind several rich veins of research including API search~\cite{Stylos:2006:MWT:1174509.1174678}, API usage pattern mining~\cite{zhong2009mapo}, API example synthesis~\cite{buse2012synthesizing}, and API documentation generation~\cite{mcmillan2011portfolio}.  The problem has long been a focus of empirical studies, which conclude that programmers often struggle to find the components they need when using APIs, and use a variety of techniques and tool support to find relevant API functions~\cite{robillard2011field,uddin2015api, meng2018application, meng2019developers,gao2020exploring}.

A scientific consensus is now emerging around \emph{interactive} API search.  The idea is that tool support should mimic the experience of asking for help from a fellow human programmer -- tool support should ask clarification questions, provide feedback, and engage in other follow-up conversation to help resolve the programmer's search task.  This consensus is evident in recent NSF and industry-sponsored workshops~\cite{dlse2019} and is articulated by fourteen leading researchers who ``advocate for a new vision for satisfying the information needs of developers''~\cite{robillard2017demand}, which they call ``on-demand developer documentation.''  The idea is that we as a research field should move towards enabling intelligent machine responses to programmer information needs.

One major barrier to interactive API search is \emph{dialogue management}.  A dialogue manager is the component of an interactive dialogue system that determines what the system should say.  At a very high level, any dialogue system will have three parts: natural language understanding (NLU), natural language generation (NLG), and a dialogue manager (DM)~\cite{burgan2016dialogue}.  The NLU component is responsible for interpreting the programmer's information need.  The NLG component is responsible for creating a natural-language reply to that need.  The DM component is responsible for telling the NLG unit what kind of reply it should generate based on the input from the NLU component, previous dialogue history, and other factors. In short, the dialogue manager is what makes \emph{interactive} search possible.  Significant research in software engineering has focused on the NLU and NLG components.  At present, dialogue management has tended to be overlooked.

In this paper, we present a dialogue manager for interactive API search. Our approach is built around a typical API search engine, which takes in a user query and returns a ranked list of relevant components. Whereas a basic search engine invariably presents these results to the user, our dialogue manager considers its confidence in the search results and the conversation history to select actions from a set of API search activities, including API recommendation and query refinement. The DM's goal is to respond to user input with an action that will most efficiently guide the user to an API function that will satisfy their functionality requirements.

We created two versions of our dialogue manager; one version follows a dialogue policy we crafted by hand for interactive API search, and the other follows a policy we optimized via reinforcement learning through interactions with a user simulator. We evaluated both versions in a synthetic evaluation with the user simulator and found that both outperformed a baseline designed to mimic a typical code search engine. We also performed a human evaluation with real programmers and found that there were advantages and disadvantages to the learned and baseline policies. 

To promote reproducibility, we release our code, dataset, and other details via an online appendix (see Section~\ref{sec:reproducibility}).



\section{Background \& Related Work}

In this section, we discuss key background technologies in interactive dialogue systems and API search.  We also introduce related work in dialogue systems for software engineering.


\subsection{Dialogue Acts}

The fundamental component of a conversation, as far as interactive dialogue systems are concerned, is the ``dialogue act.''  A dialogue act is an intermediate representation of an utterance in a conversation.  For example, a system to answer questions about the weather might respond to dialogue acts related to different types of weather events: the system may recognize a {\small \texttt{preciptationAmountQuestion}} like ``how much will it rain today?'' or a {\small \texttt{preciptationTypeQuestion}} such as ``will it rain or snow?''  Then, the system may respond with a {\small \texttt{preciptationAmountMessage}} or {\small \texttt{preciptationTypeMessage}} to answer the question.

The set of dialogue acts that an interactive dialogue system recognizes is called the system's ``action space.''  The action space differs depending on the domain.  The dialogue acts about weather are only relevant to a Q/A system about weather.  Usually the first efforts to create a dialogue system for a new domain center around defining the action space by defining the dialogue acts which the system must understand and generate.  An example within Software Engineering research is Wood~\emph{et al.}~\cite{wood2018detecting}, which identified 26 dialogue act types for a hypothetical system to help programmers debug.  This paper focuses on API search, and is based on the action space identified by Eberhart~\emph{et al.}~\cite{eberhart2020wizard}.  That paper presented a Wizard-of-Oz (WoZ) study in which programmers interacted with a hypothetical dialogue system for API usability.  We focus specifically on the API search task within the general problem of API help, and derive an action space from the dialogue acts that they identified for API search.  We provide more details of this action space in Section~\ref{sec:actionspace}.


\subsection{Task-oriented Dialogue Systems}
\label{sec:taskdialogue}

\begin{figure}[b]
    \centering
    \vspace{-.3cm}
    \includegraphics[width=6cm]{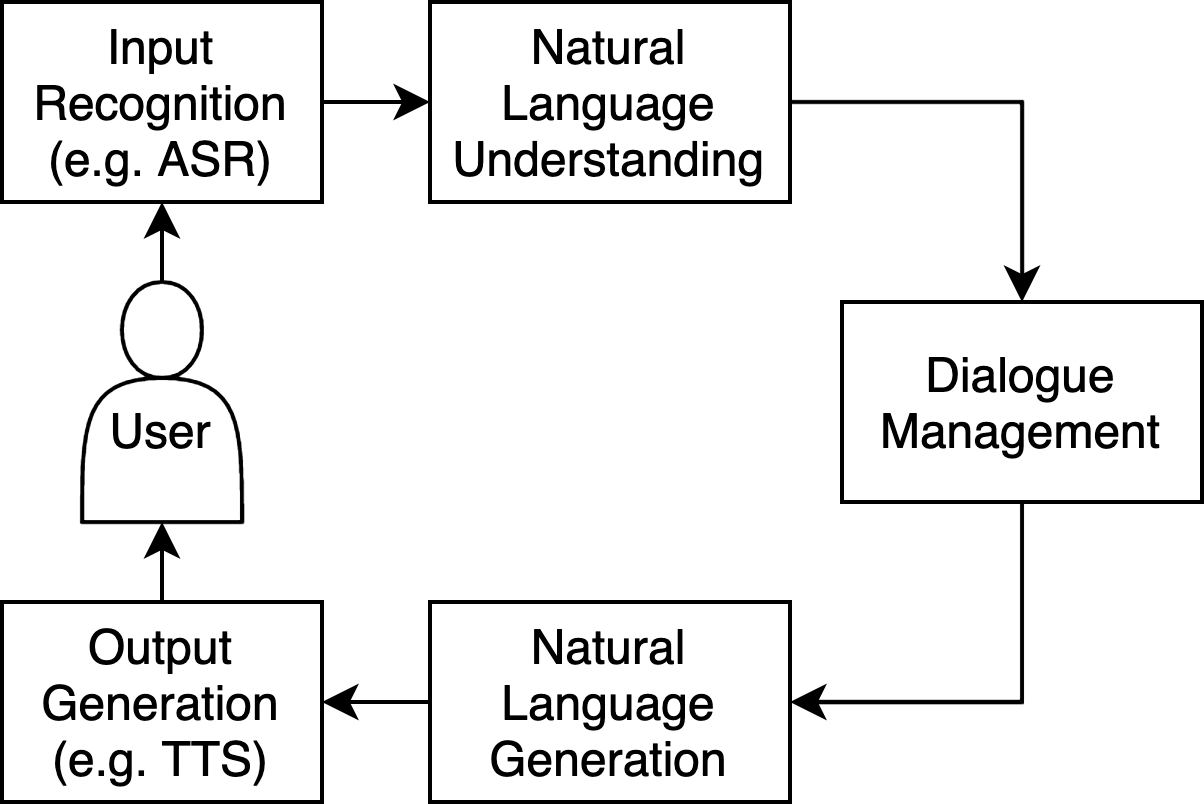}
   \vspace{-.2cm}
    \caption{A typical five-stage pipeline for task-oriented dialogue systems.}
    \label{fig:pipeline}
\end{figure} 

Task-oriented dialogue systems are software systems that help users complete tasks by communicating with them via natural language. Examples include virtual assistants (e.g. Siri, Alexa, Google Assistant), customer support bots, and automated booking systems. In contrast to general purpose ``chatbots,'' task-oriented systems are designed to help users accomplish specific goals, such as question answering~\cite{chan2019question}, data manipulation~\cite{bradley2018context}, and conversational search~\cite{zhang2018towards}. By mimicking interactions between humans, dialogue systems can provide advantages over non-conversational systems with respect to task performance and user-friendliness~\cite{rieser2011reinforcement, aggarwal2018improving}.

Essentially, what a task-oriented dialogue system does is navigate through an action space until a particular goal is accomplished.  Burgan~\cite{burgan2016dialogue} defines this navigation in terms of a five-stage pipeline, which we show in Figure~\ref{fig:pipeline}.  At a very high level, there is an Input Recognition component that converts the user's input (e.g. from speech or keyboard) into a string of words.  Then there is a Natural Language Understanding (NLU) component that decides the dialogue act of the string of words.  The NLU component also extracts information relevant to that dialogue act, such as if a question is about the weather today, tomorrow, or next week.  Then it is the Dialogue Manager's job to decide which dialogue act to use in the response, and to extract information relevant to that dialogue act.  A simple example is to use a {\small \texttt{preciptationAmountMessage}} to respond to a {\small \texttt{preciptationAmountQuestion}}, and to extract the prediction for rain amount today for that dialogue act.  Finally, a Natural Language Generation (NLG) component converts the dialogue act into a string of words that contains the relevant information, which is presented to the user via an Output Generation interface.


Note that this pipeline is in contrast to end-to-end approaches which attempt to learn NLU, NLG, and DM behaviors purely from big data input, e.g. via an encoder-decoder neural architecture~\cite{wen2016network, liu2018dialogue}.  For a time, end-to-end systems were viewed as a remedy to the significant manual effort required to build each component separately and write rules or train models at each stage~\cite{chen2017survey}.  End-to-end systems have been shown successful for tasks such as answering programmer ``basic'' questions about code, e.g. what the name or return type of a function is~\cite{bansal2021neural}.  However, recently there has been a resurgence of interest in decoupling dialogue policy and language understanding/generation in end-to-end systems, to make them more like the traditional pipeline-based architectures~\cite{he2018decoupling, burtsev2018deeppavlov, papangelis2020plato, finch2020emora}.



\subsection{Dialogue Management in Task-Oriented Systems}
\label{sec:dmtos}

The dialogue manager's job, in a nutshell, is to 1) decide which dialogue act to use in response, and 2) extract information relevant to that dialogue act sufficient for the NLG component to create a natural language text rendering of the dialogue act.  The body of rules which the manager uses to make decisions is called the \textbf{dialogue policy}.  There are three types of policy:

\begin{labeling}{a}
	\item[\emph{Single-turn}] policies provide one response to one utterance.  Q/A systems and search engines follow these policies by providing one answer to one query.
	\vspace{0.1cm}
	\item[\emph{Hand-crafted}] policies depend on rules manually written by human experts.  These policies are time consuming to make, but are common when the action space is small.
	\vspace{0.1cm}
	\item[\emph{Learned}] policies are optimized through interaction with actual or simulated users e.g., via reinforcement learning.
\end{labeling}

\subsection{API Search}






This paper focuses on dialogue management in the context of API search.  API search is the task of finding API components that fulfill a user's information need (typically by implementing some particular functionality). API search is closely related to the fields of API discovery and recommendation, code search, and software component retrieval, though we draw several distinctions.  Unlike code search~\cite{reiss2009semantics}, systems for API search do not necessarily have access to source code repositories, and may rely solely upon API documentation or specifications. API search also targets resources and operations with specific properties within APIs, unlike API discovery/recommendation and software component retrieval, which may target other software units~\cite{bawa2016algorithmic, li2014novel} or base results on code context rather than queries~\cite{liu2018effective}. Still, there is significant overlap between the research fields, and concepts and techniques from one can often be applied in the others.

API search is a well-studied field in SE research.  Examples include concept-based approaches that rely on text or other code artifacts to locate relevant API components~\cite{campbell2017nlp2code, Stylos:2006:MWT:1174509.1174678}, specification-based techniques that match user-defined constraints to clearly-defined properties of API components~\cite{zaremski1995signature, treude2015tasknav}, and structure-based approaches that rely on relationships among API components such as function calls~\cite{de2013multi, mcmillan2011portfolio, eisenberg2010apatite}.

\subsection{Interactive Dialogue in SE}


Interactive dialogue is a growing area within SE research.  Implementations of dialogue systems for SE tasks include WhyLine~\cite{ko2004designing}, TiQi~\cite{pruski2015tiqi}, Devy~\cite{bradley2018context}, and a system for generating GUIs from natural language requirements~\cite{kolthoff2019automatic}. Two recent conversational tools that are relevant to the problem of API search are OpenAPI bot and Chatbot4QR. OpenAPI bot~\cite{edopenapi} provides a natural language interface for users to query OpenAPI specifications~\cite{openapi}.  It enables users to make requests like ``show me the list of paths in the API.''  It relies on a hand-crafted dialogue policy, though it does not allow users to search for relevant API components by describing their desired functionality. In contrast, Chatbot4QR~\cite{zhang2020chatbot4qr} is a search tool for StackOverflow posts that uses heuristics to generate clarification questions.

\section{Approach}
\label{sec:approach}

Our approach consists of a dialogue manager for conversational API search.  Recall from Section~\ref{sec:dmtos} that the dialogue manager's job is to help the system navigate through the action space of dialogue acts, by selecting which dialogue act the system will use, and extracting information relevant to that dialogue act.  In this section, we 1) define the action space of dialogue acts, and 2) describe how we extract the relevant information.  Note that we describe how we create different dialogue policies as a separate section (Section~\ref{sec:policies}).  The reason is that we have two versions of our dialogue manager, and they are identical except for the dialogue policy.  One version uses a hand-crafted policy and another uses a learned policy.  Note we also use a third, single-turn policy, but only as a baseline in our evaluation in later sections.

\subsection{Action Space}
\label{sec:actionspace}

    \begin{table}[b!]
    \vspace{-0.4cm}
    \centering
     \caption{System and user action spaces.}
     \label{tab:apizarlactions}
    \renewcommand{\arraystretch}{1.1}
    \begin{tabular}{lll}
    Speaker                 & Search Function                      & Dialogue Act        \\ \hline
    \multirow{10}{*}{User}  & \multirow{2}{*}{Semantic Search}     & \texttt{provideQuery}                   \\
                            &                                      & \texttt{provideKeyword}                 \\ \cline{2-3} 
                            & \multirow{3}{*}{User Critique}       & \texttt{rejectKeywords}                 \\
                            &                                      & \texttt{rejectComponents}               \\
                            &                                      & \texttt{unsure}                          \\ \cline{2-3} 
                            & \multirow{5}{*}{Standard Navigation} & \texttt{elicitInfoAPI}                   \\
                            &                                      & \texttt{elicitInfoAllAPI}               \\
                            &                                      & \texttt{elicitSuggAPI}               \\
                            &                                      & \texttt{elicitListResults}                     \\
                            &                                      & \texttt{changePage}                     \\ \cline{2-3} 
                            & General                                 & \texttt{END}                             \\ \hline
    \multirow{8}{*}{System} & \multirow{2}{*}{Query Refinement}    & \texttt{requestQuery}             \\
                            &                                      & \texttt{suggKeywords}                \\ \cline{2-3} 
                            & \multirow{2}{*}{API Recommendation}  & \texttt{suggAPI}                       \\
                            &                                      & \texttt{suggInfoAPI}                \\ \cline{2-3} 
                            & \multirow{3}{*}{Standard Navigation} & \texttt{infoAPI}                          \\
                            &                                      & \texttt{infoAllAPI}                            \\
                            &                                      & \texttt{listResults}                            \\
                            &                                      & \texttt{changePage}                     \\ \cline{2-3} 
                            & General                                 & \texttt{START}                          
    \end{tabular}
    \end{table}

	We created an action space for both system and user dialogue acts. This is the action space within which our dialogue manager can operate.  We started with the dataset provided by Eberhart~\emph{et al.}~\cite{eberhart2020automatically} of API help dialogues, then refined the set of possible dialogue acts based on other related literature.  We show the action space in Table~\ref{tab:apizarlactions}.  We separate the actions into six categories describing different search behaviors:

    \emph{1. Semantic search}. These actions enable users to search for API components by providing unstructured natural language queries, as well as specific keywords. Duala-Ekoko and Robillard~\cite{Duala-Ekoko:2012:AAQ:2337223.2337255}, Sadowski et al.~\cite{sadowski2015developers}, and Eberhart et al.~\cite{eberhart2020wizard} have demonstrated that natural language queries play a more significant role in the API information-seeking process than e.g., signature-matching and faceted search. Furthermore, a broad range of techniques have been investigated supporting semantic search for APIs and source code~\cite{ma2018web, husain2019codesearchnet}.  

    \emph{2. User critique}. These actions enable users to reject suggestions provided by the system, or indicate that they are unsure of a suggestion's relevance. The ability for users to ``critique'' system suggestions is a pillar of the conversational search framework outlined by Radlinski and Craswell~\cite{radlinski2017theoretical}.

    \emph{3. Query refinement}. These actions enable the system to help users improve their queries by suggesting keywords that may be relevant or prompting users to reword their queries. Programmers are not always able to sufficiently articulate their information needs~\cite{Duala-Ekoko:2012:AAQ:2337223.2337255,nie2016query,zhang2020chatbot4qr,piccioni2013empirical}. Robillard et al.~\cite{robillard2017demand} explain that a system for on-demand developer documentation should help programmers improve their queries. 

    \emph{4. API Recommendation}. These actions enable the system to recommend individual API components and present corresponding documentation. API documentation structured as fragmented lists without a ``coherent, linear'' presentation can be overwhelming~\cite{robillard2011field}, and the ability to present a single item at a time can enable more targeted responses~\cite{radlinski2017theoretical,rieser2011reinforcement}. 

    \emph{5. Standard navigation}. These actions serve as basic tools for navigating results and documentation; e.g., the users should be able to explicitly request that the system present a list of results, show the next page of a list of results, or present documentation for particular components, and the system should be able to perform each of the corresponding actions. 

    \emph{6. General}. Finally, the system and the users have actions to start and end the dialogue, respectively.

    We limit the number of components or keywords that the system can present in a given turn, and we only permit the user to present one query or keyword or request information about one function at a time. Furthermore, we identify specific properties of API components that can be requested or shared; we discuss these in the following section.


\subsection{Knowledge Management}

	Knowledge management is the part of the dialogue manager that extracts information relevant to different dialogue acts, so that the NLG component can render a text response.  We implement it as a dataset that provides operations for the system to query and retrieve different types of information. 

    \subsubsection{API Dataset}
    \label{sec:apizarlds}
    The API dataset consists of reference documentation for components in an API, indexed by components' fully-qualified names. We narrow our API search task and increase the generalizibility of our approach by targeting only API components and including only documentation resources that are available for the vast majority of APIs, regardless of domain, library language, popularity, or other factors (in other words, the information that would be available in a typical API definition).  To that end, we associate each API component with 1) its signature, 2) its summary description, and 3) other documented properties. 

    A component's signature comprises its name, return type, parameter names, and parameter types. Its summary description is the human-written (or in some cases, automatically-generated) text explanation of the component's primary purpose/functionality. At minimum, most common documentation formats call for developers to include a summary description. Finally, functions can be associated with any number of other documented properties, including longer descriptions, details about their parameters and return values, usage examples, related components, categories, class hierarchies, and more. These properties are stored as text strings in the dataset. During dialogues, users can request any or all information associated with a component, and the dialogue system can retrieve them. 

    In order to enable simple semantic search, we associate each API component with a TF-IDF-derived \emph{search vector.}' TF-IDF (term frequency-inverse document frequency) vectors represent how relevant individual terms are to a text document by considering word frequency within a document and the number of documents in the dataset that contain each word. In information retrieval, documents are represented by TF-IDF vectors to enable a simple form of semantic search. In the API dataset, each component's search vector is generated by calculating TF-IDF vectors for each of its associated properties, and then averaging the signature vector, the summary description vector, and a third vector that is itself an average of the TF-IDF vectors of that component's other properties. This representation is inspired by Yu et al.~\cite{yu2016apibook}, who similarly weighted TF-IDF vectors of component properties in order to enhance the contributions of keywords in the components' names and descriptions over those keywords in other properties.

    \subsubsection{Component Search}
    When users modify their search by providing a query, providing keywords, or rejecting components, a \emph{similarity score} $s$ between 0 and 1 is calculated for each component in the dataset. The similarity score is calculated in two steps: first, if the user has provided a query, it is transformed into a TF-IDF vector, and each API component in the dataset is scored by calculating the cosine similarity between it's search vector and the query vector. Second, any provided keywords and rejected components are applied as binary filters. The components are then ranked by their similarity scores (with ties decided randomly). 

    When the system selects the {\small \texttt{listResults}} action, it simply retrieves the first $N$ results from the ranked list (where $N$ is a predefined parameter of the search environment). When the system selects the {\small \texttt{changePage}} action, it increments a result index $r$ by $N$ and retrieves the next $N$ results. When it selects the {\small \texttt{suggAPI}} or {\small \texttt{suggInfoAPI}} action, it retrieves the result at index $r$ and then increments $r$ by 1. The dialogue manager resets $r$ to 0 whenever the system selects the {\small \texttt{listResults}} action or the user's search is modified.

    \subsubsection{Keyword recommendation}
    When the system selects the {\small \texttt{suggKeywords}} action, it uses a naive keyword recommendation approach to retrieve the top $K$ keywords that are potentially relevant to the user's search (where $K$ is a predefined parameter of the search environment). First, it averages the TF-IDF search vectors of the 20 components with the highest similarity rankings. Then, it sets the indices of all terms that appear in the provided or rejected keyword lists or the user's query to 0. Finally, it returns the keywords corresponding to the indices of the $K$ largest values in the vector.

\subsection{Dialogue State Tracking}
\label{sec:statespace}

	The dialogue manager also keeps track of the dialogue state, which is just a record of the conversation so far.   In line with Aggarwal et al.\cite{aggarwal2018improving} and Rieser and Lemon~\cite{rieser2011reinforcement}, we define the state to consist of the following information: the most recent system dialogue act type, the most recent user dialogue act type, the dialogue length (measured in turns), and the similarity scores of the API components. The dialogue manager updates these state values whenever the DM receives a new user dialogue act or produces a dialogue act response.

\section{Dialogue Policies}
\label{sec:policies}

Recall from Section~\ref{sec:dmtos} that there are three types of dialogue policy: 1) single-turn, 2) hand-crafted, and 3) learned.  We explore all three in this paper.  We consider the hand-crafted and learned policies to be novel contributions of this paper.  However, the single-turn policy is a baseline from source code search and Q/A dialogue, which we use only for comparison in our evaluations.

The key element to our hand-crafted and learned policies is the \textit{reward function}.  This function defines how well the system performed in the conversation.  For the hand-crafted policy, we optimized for this function by manually writing rules to define how the system behaves in different situations.  For the learned policy, we trained a reinforcement learning model to optimize for this reward function.


\subsection{Reward Function}
\label{sec:reward}

We implement a reward function that prioritizes dialogue length and concise system dialogue acts, in line with reward functions for conversational search used by Aggarwal et al.~\cite{aggarwal2018improving} and Rieser and Lemon ~\cite{rieser2011reinforcement}. Specifically, the system incurs a penalty of $-1$ each turn, incentivizing it to help the user complete his or her search task, and to do so as quickly as possible. There are also penalties for system acts that present users with lists of search results or entire pages of documentation: a $-.3$ penalty for {\small \texttt{listResults}} and {\small \texttt{changePage}} acts, and a $-.5$ penalty for {\small \texttt{infoAllAPI}} and {\small \texttt{suggInfoAllAPI}}. These penalties are not applied when the system is responding to a corresponding user ``'standard navigation'' act (e.g., {\small \texttt{listResults}} in response to  {\small \texttt{elicitListResults}}). We refer to these penalties as the ``core'' reward function, which we use to evaluate the different dialogue policies: 

\vspace{-.5cm}
\begin{equation}
\begin{multlined}R_{Core}(turn) = -1 - r_{DialogueActPenalty}(turn)$$
\end{multlined}
\end{equation}
\vspace{-.5cm}

We use a modified version of the reward function to train the learned policy. The modified reward function is designed to speed up the training process by rewarding progress toward the search goal and penalizing certain incorrect behaviors. Specifically, we define an extrinsic reward that the system earns when the ranking of the user's target function improves in the search results (the reward is proportional to the rank improvement, with a maximum value of $+5$). When the system successfully completes the search task, it earns a reward of $+10$. We also harshly penalize the system ($-10$) when it does not respond to users' ``standard navigation'' acts with the corresponding system act. Finally, the system incurs a small penalty ($-1$) when the user selects the {\small \texttt{unsure}} act. 

\vspace{-.5cm}
\begin{equation}
\begin{multlined}
$$R_{Training}(turn) = R_{Core}(turn) \\
+ r_{TrainingPenalties}(turn) + r_{TrainingRewards}(turn)\\
\end{multlined}
\end{equation}
\vspace{-.9cm}

\subsection{Hand-crafted Policy}

Our hand-crafted policy includes several rules and parameterized thresholds. First, the hand-crafted policy responds to all ``standard navigation'' user actions with the corresponding system actions, and it responds to the {\small \texttt{unsure}} dialogue act by simply listing the current search results.  For all other user action types, it considers the the similarity score $s$ of the top-ranked component in the API dataset. If the user has not recently selected the {\small \texttt{unsure}} act, the policy first checks whether $s$ is below either of two thresholds indicating that it should attempt to refine the user's query by requesting that the user elaborate, or by suggesting potential keywords. Otherwise, the policy checks if $s$ is above either of two thresholds indicating that it has confidence to recommend the function and it's corresponding documentation, or the function alone. If the similarity score is lower than these thresholds, the policy will list the top-$N$ search results. The exact values for the hard-coded thresholds are chosen by performing a grid search, evaluating the policies against the reward function using a user simulator (described in the following section).

The advantage of this policy is that it encodes a domain expert's intuition about how the system should behave, so is likely to be effective in the situations for which it is designed.  The disadvantage is that a hand-crafted policy is expensive to create, and is limited by the designer's understanding of the search task.  The policy is also inflexible; if a change to the action space is desired, new rules must be added.

\subsection{Learned Policy}

We use Deep Q-learning to train a dialogue policy.  Deep Q-learning is a reinforcement learning algorithm that is recommended by Cuayáhuitl~\cite{cuayahuitl2017simpleds} for creating dialogue policies.  The algorithm trains a neural network to estimate how effective different dialogue acts are in different dialogue states by repeatedly experimenting with policies in a simulated environment. Mnih~\emph{et al.}~\cite{mnih2013playing, mnih2015human} provide further background.

We formulate the learning problem as a partially observable Markov decision process (POMDP), defining four components: an action space, a state space, a reward function, and an environment.   The action space is the space we defined in Section~\ref{sec:actionspace}.  The reward function is the function we defined in Section~\ref{sec:reward}.  The state space is the state information we maintain in Section~\ref{sec:statespace}. The environment is provided by a User Simulator of our own design, which we describe in the next paragraphs.  Note that there are many implementation details which far exceed the amount of space available in this paper, so we make our complete implementation and other data available via our online repository (see ~\Cref{sec:reproducibility}).





\subsubsection{User Simulation}
\label{sec:usersimulator}

We created a user simulator to serve as the environment in which our dialogue manager is trained.  The user simulator is itself a simple dialogue manager, accepting a dialogue act from the API search dialogue manager and returning an appropriate dialogue act in response. Note that, while the API search dialogue policy is optimized for a particular reward function, the user simulator is not intended to behave ``optimally'' -- on the contrary, it is intended to simulate a range of  behaviors which may or may not actually comprise efficient information-seeking strategies. The idea is to simulate ``real'' user behavior, rather than ``ideal'' user behavior.

We adapt an agenda-based approach to user simulation. Agenda-based user simulation~\cite{schatzmann2007agenda} involves modeling a user's goals and an ``agenda'' -- the actions the user intends to take. The goals are subdivided into constraints (i.e., what information the user can and cannot provide) and requests (the user's information need); over the course of a dialogue, the constraints and requests are updated as the user collects new information. The agenda is typically structured as a stack; however, we make modifications for our particular search task in order to better capture the API browsing process described by Kelleher and Inchinco~\cite{kelleher2019towards} and Eberhart et al.~\cite{eberhart2020wizard}. 

At the start of each dialogue, the user simulator randomly selects a function from the dataset to serve as the information target, but the identify of the function itself is hidden. Instead, the simulator generates a query for the function by extracting terms from the function's dataset entry. It uses the function's TF-IDF search vector to set the likelihood of selecting different terms, and it incorporates a query-error parameter (randomly set at the start of the dialogue) that increases the likelihood of all terms (allowing for incorrect terms to appear in the query). The resultant query is the user's initial ``constraint.''

The simulator's ``requests'' comprise a list of candidate functions that the user would like to learn about. Whenever the search system lists API functions or recommends a function, each function has a random chance to be added to the candidates list. Candidates are associated with an ``evidence'' value; when the system informs the user about any property of a candidate function, the evidence increases. When a candidate's evidence reaches a threshold value, it is removed from the candidate list; if it was the target function, the user is prompted to end the conversation. Otherwise, the simulator removes the function from the pool of potential candidates and decreases its query-error parameter by a small amount to model the ``learning'' process users undergo during search tasks~\cite{kelleher2019towards}.

The simulator also maintains a variable indicating the user's ability provide a new query or keywords. Whenever the user provides this information, the variable decreases; as the user is exposed to more API functions and documentation, the variable increases. If the variable is below a corresponding threshold when the user attempts to provide a query or keyword, the user instead selects the {\small \texttt{unsure}} act.

The user simulator processes incoming system dialogue acts by first processing new candidates or evidence, and then identifying the list of available actions by considering whether the task has been completed and whether there are candidate functions to inquire about. Finally, the simulator selects an action from the list using a bigram probability model (i.e., one that returns the probability of a user dialogue act type in response to the most recent system dialogue act type) derived from the Wizard of Oz data collected by Eberhart et al.~\cite{eberhart2020wizard}.

\subsection{Single-turn, top-$N$ Policy}

The top-$N$ policy emulates behavior similar to that exhibited by the simple search utilities that are integrated in some HTML documentation formats, such as Doxygen~\cite{DoxygenWebsite}. Like the other two policies, this policy responds to user's standard navigation requests with the corresponding system action. All other user actions prompt this policy to list the top-$N$ search results (by way of the {\small \texttt{listResults}} dialogue act). Whereas the other two are capable of mixed-initiative interaction, this policy is purely reactive, unable to suggest query refinements or proactively recommend API functions.  This policy is the most simple to implement, and it should align most closely with users' expectations for and past experience with search utilities. However, the absence of features that are more characteristic of interactive documentation may increase the difficulty of the search task relative to the other policies.

\section{Synthetic Evaluation}
\label{sec:quanteval}

\begin{figure*}
    \centering
    \begin{tabular}{cc}
        \includegraphics[width=75mm]{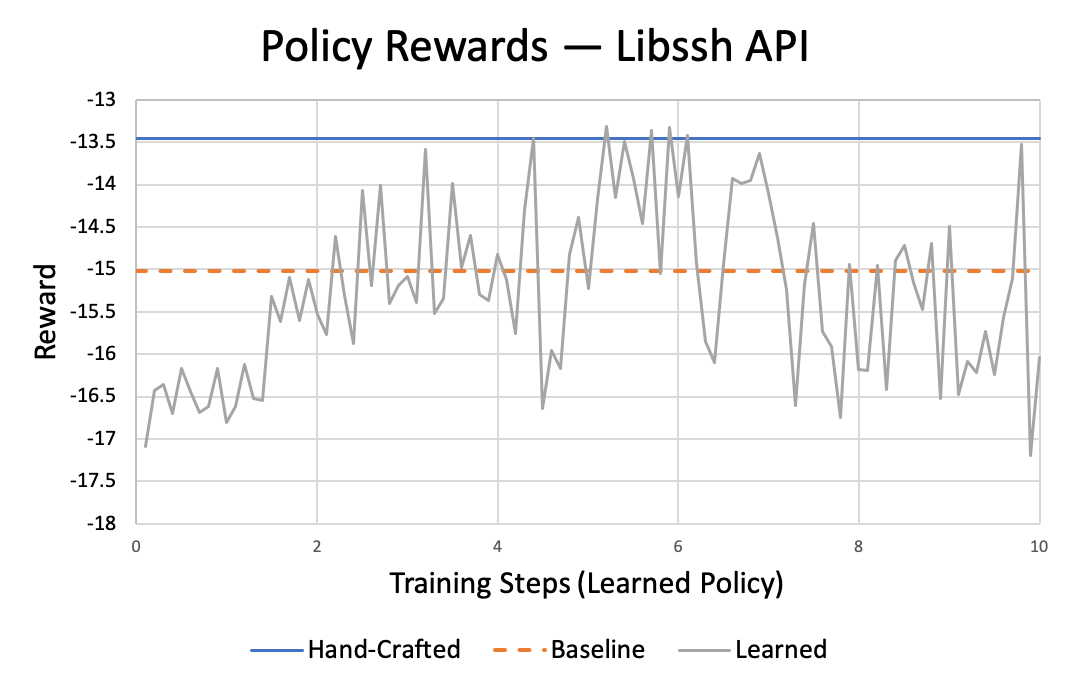} &   \includegraphics[width=75mm]{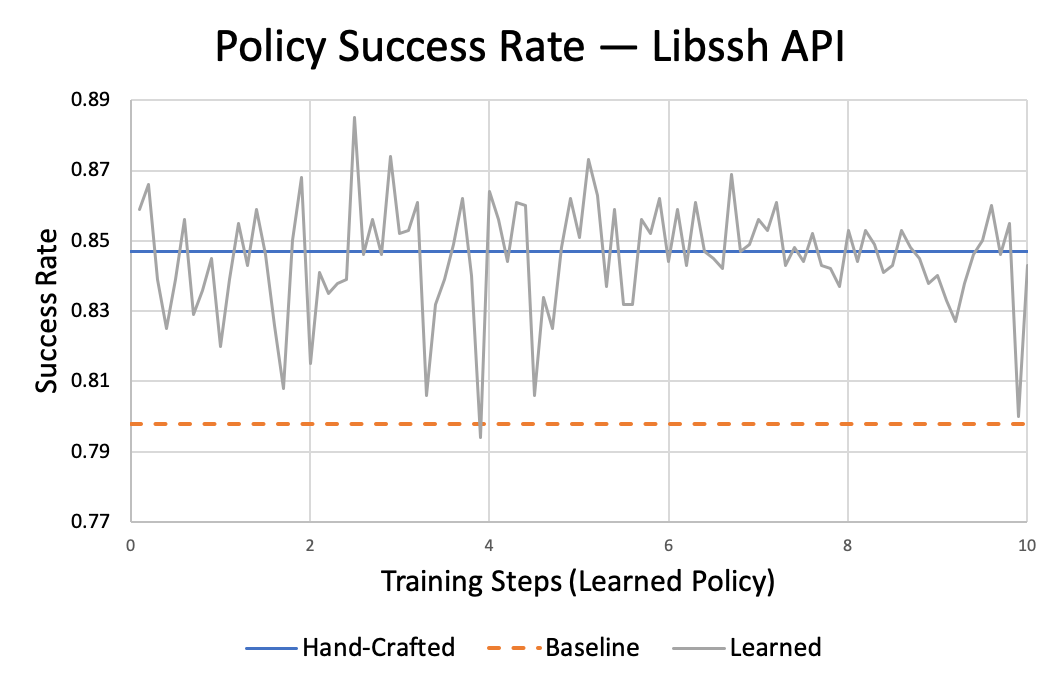} \\
        \includegraphics[width=75mm]{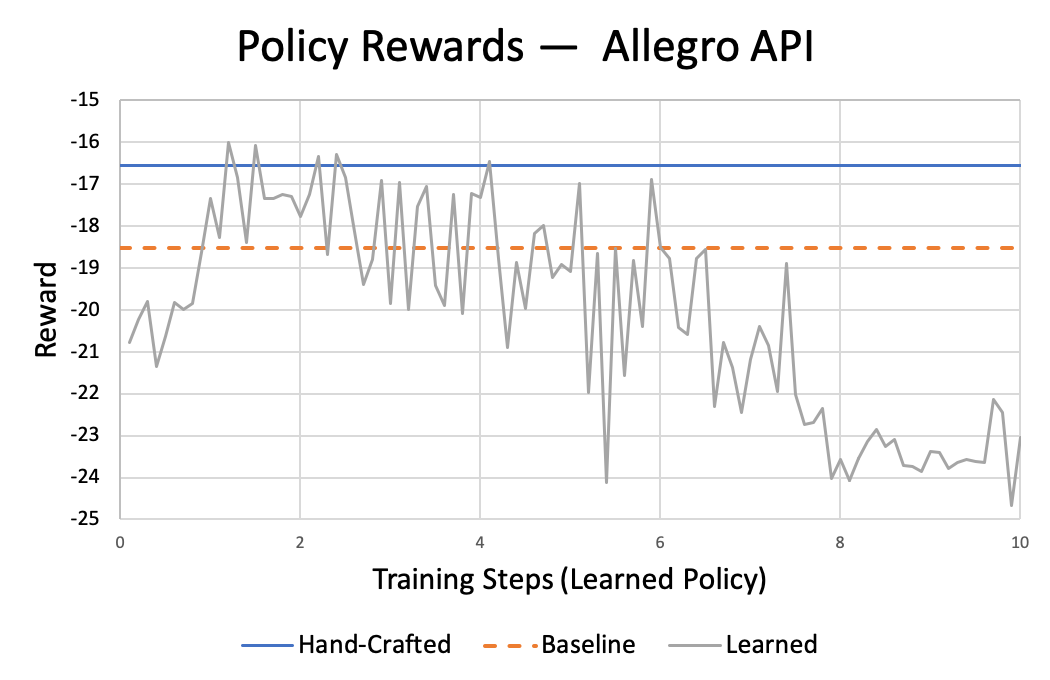} &   \includegraphics[width=75mm]{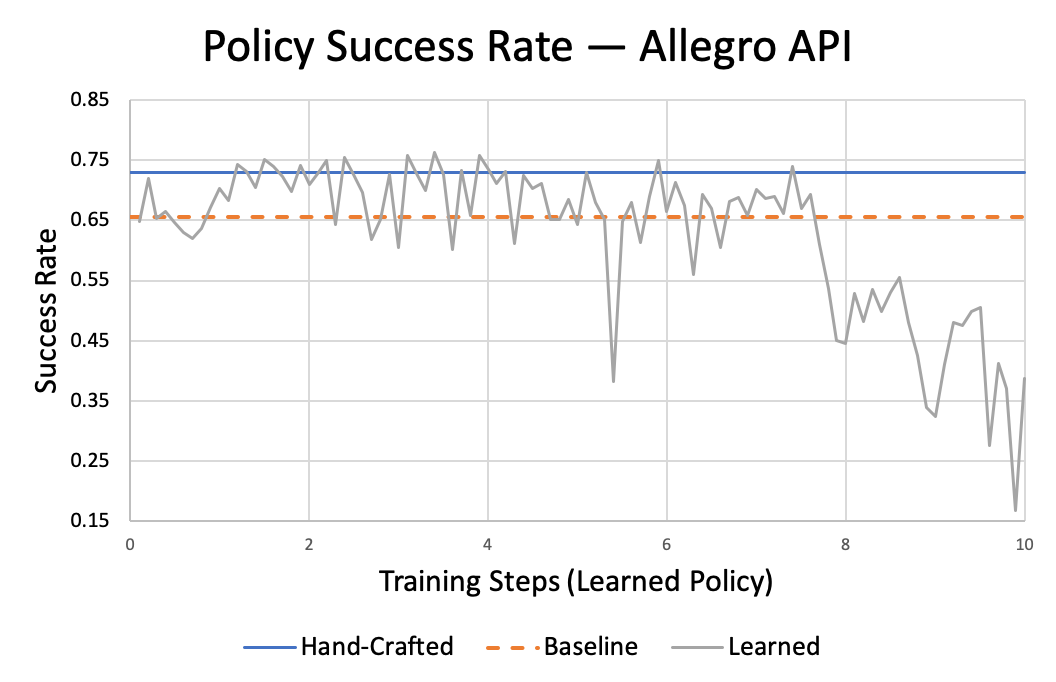} \\
    \end{tabular}
    \vspace{-0.2cm}
    \caption{Line graphs of system performance during our synthetic study.  The blue line indicates average hand-crafted performance, e.g., -13.5 for {\small \texttt{libssh}} reward.  The dashed orange line indicates average single-turn baseline policy performance.  The gray line is the performance of the learned model at that training step. The x-axis indicates millions of training steps for the learned policy. Note peak performance for {\small \texttt{libssh}} around five million training steps.}
	\label{fig:quantresults}
	\vspace{-0.4cm}
\end{figure*}

We perform a synthetic and a human evaluation in this paper.  This section describes the synthetic evaluation.  The purpose is to measure the performance of our dialogue manager in terms of the reward function we defined.  We compare versions of the dialogue manager using the three policies from the previous section: single-turn, hand-crafted, and learned.  To that end, we ask the following research question ($RQ$):

\smallskip
\begin{description}
\item[$RQ_1$] How well does the dialogue manager perform in simulated dialogues using the hand-crafted policy, the learned policy, and the baseline single-turn policy?
\end{description}
\smallskip

The purpose of $RQ_1$ is to discover whether interactive dialogue policies can improve API search by reducing the number of turns and the amount of information required to help users find useful API components. We optimize policies for the reward function (either via a learned or hand-crafted process), so we aim to measure how well the policies perform in terms of this function.  Likewise, because the user simulator can create an arbitrary number of simulated conversations, it can help provide a wide range of samples for comparison.  Also, because the conversations generated by the user simulator are only dialogue acts and related data (instead of text), $RQ_1$ allows us to compare the policies without the potentially-biasing effects of an NLU and NLG interface.

Note that this evaluation does not seek to compare individual methods for API retrieval, recommendation, or query refinement; e.g., a dialogue policy using a state-of-the-art approach for API retrieval would presumably outperform one that relies on simpler TFIDF vectors. Instead, the goal is to quantify how efficiently different dialogue policies leveraging the \emph{same} functionality can guide users through their searches.


\subsection{Methodology}
\label{sec:quantmeth}
Our methodology for answering $RQ_1$ is to generate sample conversations with the user simulator, and play out those sample conversations with our dialogue manager using different dialogue policies. For each of the three policies, we used the dialogue manager to play out 1000 conversations with the user simulator.  Then we calculated two metrics for each conversation: 1) total reward, and 2) success rate.  The total reward is the sum of the output of the reward function for each turn in the conversation.  The success rate is the percent of conversations for which the dialogue system was ``successful'' within 25 turns.  Recall that the user simulator has a hidden variable representing the function for which it is lookingl the dialogue system is ``successful'' if it finds that function. We leave out metrics that are typically used to evaluate API search methods (e.g., mean reciprocal rank and discounted cumulative gain), which would measure the efficacy of the underlying search functionality rather than the dialogue policies themselves. 

For each policy evaluation, we provided the user simulator an identical sequence of 1000 random seeds that were used generate identical conversation staring points (e.g., target functions and queries) and user behavior. In other words, the $i^{th}$ evaluation dialogue always used the same set of initial user parameters, $U_i$, which were different for each dialogue in the evaluation. Two identical dialogue policies would prompt identical behavior from the user simulator, but policy differences that resulted in the system selecting a different action during some dialogue $i$ could alter the course of that dialogue.

The number of API components that the system could list with a {\small \texttt{listResults}} dialogue act ($N$) and the number of keywords that the system could recommend with a {\small \texttt{suggKeywords}} act ($K$) were both set to $6$.  We selected $6$ in light of the ``7 plus or minus 2'' rule for working memory~\cite{kareev2000seven}.

\begin{figure}[!b]
	\centering
	\vspace{-0.6cm}
	\includegraphics[width=80mm]{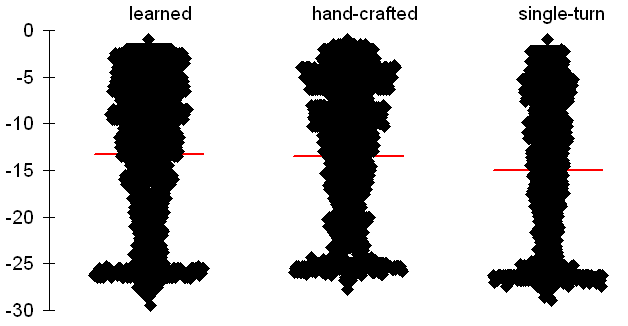}

	\vspace{0.1cm}

	\centering
	\scriptsize
	
	\begin{tabular}{ll}
		\multicolumn{2}{l}{Friedman Test Results} \\ \hline
		Q (obs)           & 50.610                \\
		Q (critical)      & 5.991                 \\
		DF                & 2                     \\
		one-tailed p      & \textless{}0.0001     \\
		alpha             & 0.05                 
	\end{tabular}

	\vspace{0.2cm}
	
	\begin{tabular}{lll}
		& learned         & hand-crafted \\ \cline{2-3} 
		\multicolumn{1}{l|}{hand-crafted} & 43.500 &              \\
		\multicolumn{1}{l|}{single-turn}  & \textbf{286.500}         & \textbf{243.000}      \\ \hline
		\multicolumn{3}{l}{critical value for difference: 87.652}         
	\end{tabular}

	\caption{The scattergrams above show the value for the reward function for the 1000 simulated conversations in our synthetic evaluation.  The tables show Friedman test results.  Bold indicates statistically-significant differences.}
    \label{fig:rfscattergrams}
\end{figure}


This evaluation used two C API datasets: {\small \texttt{libssh}}, an API for creating and using SSH network connections, and {\small \texttt{Allegro}}, an API for video game programming. The {\small \texttt{libssh}} dataset comprises 264 functions, while the {\small \texttt{Allegro}} dataset comprises 917 functions. We chose these APIs for three reasons: 1) C APIs do not include class hierarchies, enabling an emphasis on concept-based search, 2) the size and domain differences help us perceive how well the policies can generalize to a broader range of APIs, and 3) these were the APIs used in the API search experiments by Eberhart~\emph{et al.}~\cite{eberhart2020automatically}.

\vspace{-0.1cm}
\subsection{Results}

\begin{table*}[]
\centering
\caption{Actions chosen by the learned policy (vertical) and actions that the hand-crafted policy would have chosen given the same dialogue state and search results (horizontal) in the {\small \texttt{libssh}} evaluation.}

\label{tab:quantdiff}
\renewcommand{\arraystretch}{1.3}
\begin{tabular}{clrrrrrrrrr}
\multicolumn{1}{l}{}                                                                            &                                  & \multicolumn{8}{c}{Hand-Crafted Policy Action}                                                                  & \multicolumn{1}{l}{} \\
                                                                                                & \multicolumn{1}{l|}{}            & \begin{tabular}[c]{@{}r@{}}\texttt{elicit-}\\ \texttt{Query}\end{tabular} & \begin{tabular}[c]{@{}r@{}}\texttt{sugg-}\\ \texttt{Keyword}\end{tabular} & \begin{tabular}[c]{@{}r@{}}\texttt{info-}\\ \texttt{API}\end{tabular} & \begin{tabular}[c]{@{}r@{}}\texttt{infoAll-}\\ \texttt{API}\end{tabular} & \begin{tabular}[c]{@{}r@{}}\texttt{sugg-}\\ \texttt{API}\end{tabular} & \begin{tabular}[c]{@{}r@{}}\texttt{suggInfo-}\\ \texttt{API}\end{tabular} & \begin{tabular}[c]{@{}r@{}}\texttt{list-}\\ \texttt{Results}\end{tabular} & \multicolumn{1}{r|}{\begin{tabular}[c]{@{}r@{}}\texttt{change-}\\ \texttt{Page}\end{tabular}} & Total                \\ \cline{2-11} 
\multirow{8}{*}{\begin{tabular}[c]{@{}c@{}}Learned Policy \\ Action \\ (selected)\end{tabular}} & \multicolumn{1}{l|}{\texttt{elicitQuery}}    & 163      & 0           & 0    & 0       & 0       & 0           & 0           & \multicolumn{1}{r|}{0}          & 163                  \\
                                                                                                & \multicolumn{1}{l|}{\texttt{suggKeyword}} & 25       & 0           & 0    & 0       & 0       & 0           & 2           & \multicolumn{1}{r|}{0}          & 27                   \\
                                                                                                & \multicolumn{1}{l|}{\texttt{infoAPI}}        & 0        & 0           & 847  & 0       & 0       & 0           & 0           & \multicolumn{1}{r|}{0}          & 847                  \\
                                                                                                & \multicolumn{1}{l|}{\texttt{infoAllAPI}}     & 0        & 0           & 0    & 728     & 0       & 0           & 0           & \multicolumn{1}{r|}{0}          & 728                  \\
                                                                                                & \multicolumn{1}{l|}{\texttt{suggAPI}}     & 36       & 190         & 0    & 0       & 963     & 59          & 211         & \multicolumn{1}{r|}{0}          & 1459                 \\
                                                                                                & \multicolumn{1}{l|}{\texttt{suggInfoAPI}} & 10       & 11          & 0    & 0       & 584     & 30          & 333         & \multicolumn{1}{r|}{1}          & 969                  \\
                                                                                                & \multicolumn{1}{l|}{\texttt{listResults}} & 5        & 0           & 0    & 0       & 37      & 0           & 611         & \multicolumn{1}{r|}{0}          & 653                  \\
                                                                                                & \multicolumn{1}{l|}{\texttt{changePage}}  & 3        & 0           & 0    & 0       & 1       & 0           & 0           & \multicolumn{1}{r|}{249}        & 253                 
\end{tabular}
\vspace{-0.2cm}
\end{table*}

We found that the hand-crafted and learned policies outperformed the single-turn policy to a statistically-significant degree.  We also found that while the learned policy is able to slightly outperform the hand-crafted policy in several instances according to average reward, this different is not statistically significant.  Consider Figure~\ref{fig:quantresults}.  The blue line is the average reward or success rate for the hand-crafted policy.  At around five million training steps, the learned policy slightly outperforms it.  Though as Figure~\ref{fig:rfscattergrams} shows, this difference is not statistically significant.  Note the scattergrams in Figure~\ref{fig:rfscattergrams}.  The left-most plot shows values for the best-performing learned policy for {\small \texttt{libssh}}.  The mean is very slightly higher than the mean for the hand-crafted policy, and both are higher than the single-turn policy.  However, we note that these are \emph{paired} results in that each scattergram has a point related to each conversation starting point from the user simulator.  In general, for each conversation starting point, the hand-crafted and learned policies do markedly better than the single-turn policy.  The evidence for this finding is from a Friedman paired test, also shown in Figure~\ref{fig:rfscattergrams}.  The learned policy is slightly but not significantly better than the hand-crafted policy.

This result is important for two reasons.  First, it means that a high-quality policy can be learned from data.  The hand-crafted policy is the result of many hours of manual effort.  It achieves very strong performance, but at very high cost.  But we are able to match, even slightly outperform, this policy with a relatively inexpensive learning-based procedure.  As Figure~\ref{fig:quantresults} shows, slightly superior performance to the hand-crafted policy is reached after a few million training steps, which corresponds to around 48 hours of training time on our hardware (Xeon E5-1650 CPU, 128GB ram).  Another reason this result is important is that both policies clearly outperform the baseline single-turn policy, which corresponds to typical search engine-like behavior.  Note the narrower shape of the scattergram in Figure~\ref{fig:rfscattergrams} for the single-turn policy.  A major reason for this shape is the larger number of conversations at the conversation cutoff point of 25 turns, which is visible as the upside-down T shape.  These are unsuccessful conversations.  For the single-turn policy, it indicates cases where the user simulator kept refining the query over and over without finding the answer, which is a common failure case for search engines.

The average rewards achieved by the best-performing versions of the learned policy were only slightly-higher than those achieved by the hand-crafted policy; however, the two policies frequently selected different actions; in fact, the learned policy diverged from the hand-crafted policy in $80.1\%$ of the {\small \texttt{libssh}} dialogues and $92.3\%$ of the {\small \texttt{Allegro}} dialogues. Table~\ref{tab:quantdiff} show all actions selected by learned policy (y-axis) in the {\small \texttt{libssh}} evaluation and the actions that the hand-crafted policy would have selected given the same dialogue states and search results (x-axis). For example, the learned policy chose the {\small \texttt{suggAPI}} dialogue act $1459$ times in the evaluation. In 963 of those instances, the hand-crafted policy would have made the same decision, but in 211, the hand-crafted policy would have instead presented a list of results.

Generally speaking, the learned policy differed by choosing actions that indicated greater confidence in the search result(s), e.g., selecting the {\small \texttt{suggInfoAPI}} act (which presents a recommended function with all corresponding documentation) rather than the {\small \texttt{suggAPI}} act (which presents a function with only a summary description), as in the following example:


\vspace{-0.2cm}
\begin{excerpt}
\label{example}
\begin{tcolorbox}[left=-10pt,right=-20pt,top=5pt,bottom=5pt]
\begin{flushleft}
\begin{dialogue}

\speak{User} {\small \texttt{provideKeyword}}(``knownhost'')

\medskip
\direct{Turn: 9. Results: 5. Highest similarity score: .206}
\medskip
\speak{Learned Policy} {\small \texttt{suggInfoAPI}}(``ssh\_write\_knownhost'')
\speak{HC Policy} {\small \texttt{suggAPI}}(``ssh\_write\_knownhost'')
\end{dialogue}
\end{flushleft}
\end{tcolorbox}
\end{excerpt}
\vspace{-0.2cm}

In this example, the results' highest similarity score (.206) was below the threshold value required by the hand-crafted policy to select the {\small \texttt{suggInfoAPI}} act. And in this case the learned policy's decision paid off, as ``sh\_write\_knownhost'' was the user simulator's target function. Other times, the learned policy did the opposite, and chose less-committed actions despite high similarity scores in the results:

\vspace{-0.2cm}
\begin{excerpt}
\label{example}
\begin{tcolorbox}[left=-10pt,right=-20pt,top=5pt,bottom=5pt]
\begin{flushleft}
\begin{dialogue}

\speak{User} {\small \texttt{provideKeyword}}(``knownhost'')

\medskip
\direct{Turn: 15. Results: 10. Highest similarity score: .597}
\medskip

\speak{Learned Policy} {\small \texttt{suggAPI}}(``ssh\_poll\_ctx\_add'')
\speak{HC Policy} {\small \texttt{suggInfoAPI}}(``ssh\_poll\_ctx\_add'')
\end{dialogue}
\end{flushleft}
\end{tcolorbox}
\end{excerpt}
\vspace{-0.2cm}

In this case, ``ssh\_poll\_ctx\_add'' was not the user simulator's desired function, and the learned policy may have been able to avoid an additional penalty by forgoing the more costly {\small \texttt{suggInfoAPI}} act. These examples demonstrate the learned policy's ability to learn to interpret potentially obscure dialogue/search features to select efficient actions.

\subsection{Threats to Validity}
\label{sec:ttvquant}
As in any evaluation, ours carries a number of threats to validity. First, a key threat to internal validity includes the assignment of user simulator behavior parameters. It is possible that different behavior parameters could have led to dialogues in which the relative performance of different dialogue policies varied. To reduce this threat, we assigned behavior parameters randomly and evaluated policies across a relatively large number of test episodes (1000), in line with related literature~\cite{chandramohan2011user}.  Second, threats to external validity include the selection of APIs, the selection of dialogue manager parameters, and the heuristics used for user goal modeling. Results observed with the two APIs evaluated may not generalize to other APIs, particularly those in other languages/domains, larger APIs (or collections of APIs), and APIs with different structural hierarchies.  Dialogue managers implemented with different values for $K$ and $N$ may see different trends in policy efficacy, and in real search scenarios, programmers' goals may not resemble targeted searches for specific functionality, meaning that the policies evaluated may not generalize to different programmer information-seeking strategies. Third, threats to construct validity include the design of the user simulator and reward function. The simulations and metrics used to train and evaluate the policies may not accurately measure the value of a dialogue manager to actual users. To help address this threat, we perform a user study designed to illuminate the performance of the dialogue manager with real programmers.

\section{Human Evaluation}

This section describes our evaluation with four human programmers.  While the synthetic study provides a broad view based on 1000 conversation starting points, those conversations were generated with simulated users.  This study provides a more in-depth view based on a small number of conversations, but generated via interaction with actual humans performing API search tasks. We ask the following $RQ$:

\begin{description}
\item[$RQ_2$] Does the learned policy outperform the baseline single-turn policy with respect to the reward function and success rate, when interacting with humans?
\end{description}

The rationale for $RQ_2$ is that the results in the previous experiment involved synthetic conversations, and so may differ from actual human behavior. The learned policy led to the best-performing configuration of our approach in Section~\ref{sec:quanteval}, and the single-turn policy represents a typical configuration for most search tools.  We note that this evaluation is constrained in size and scope; it primarily serves as a proof-of-concept and supplements the findings from the previous section.   


\subsection{Methodology}

Our methodology for answering $RQ_2$ is to ask programmers to complete a series of API search tasks by using a dialogue system for API search that implemented our dialogue manager. We recorded each user's interactions with the system and their answers to the search tasks.  We measured the reward for the conversations using the same reward function as the previous evaluation.  We measured success rate based on the user's task feedback. We had a reference API component for each search task (though the user did not know this component, we only provided him or her with the task); if the user found the component with the search tool and entered it as an answer to the search task (indicating that he or she not only found the answer, but recognized it as the answer) within 25 turns, we considered the conversation to be ``successful.''

\subsubsection{Search Tool}

\begin{figure}[t]
    \centering
    \vspace{-0.15cm}
    \includegraphics[width=6cm]{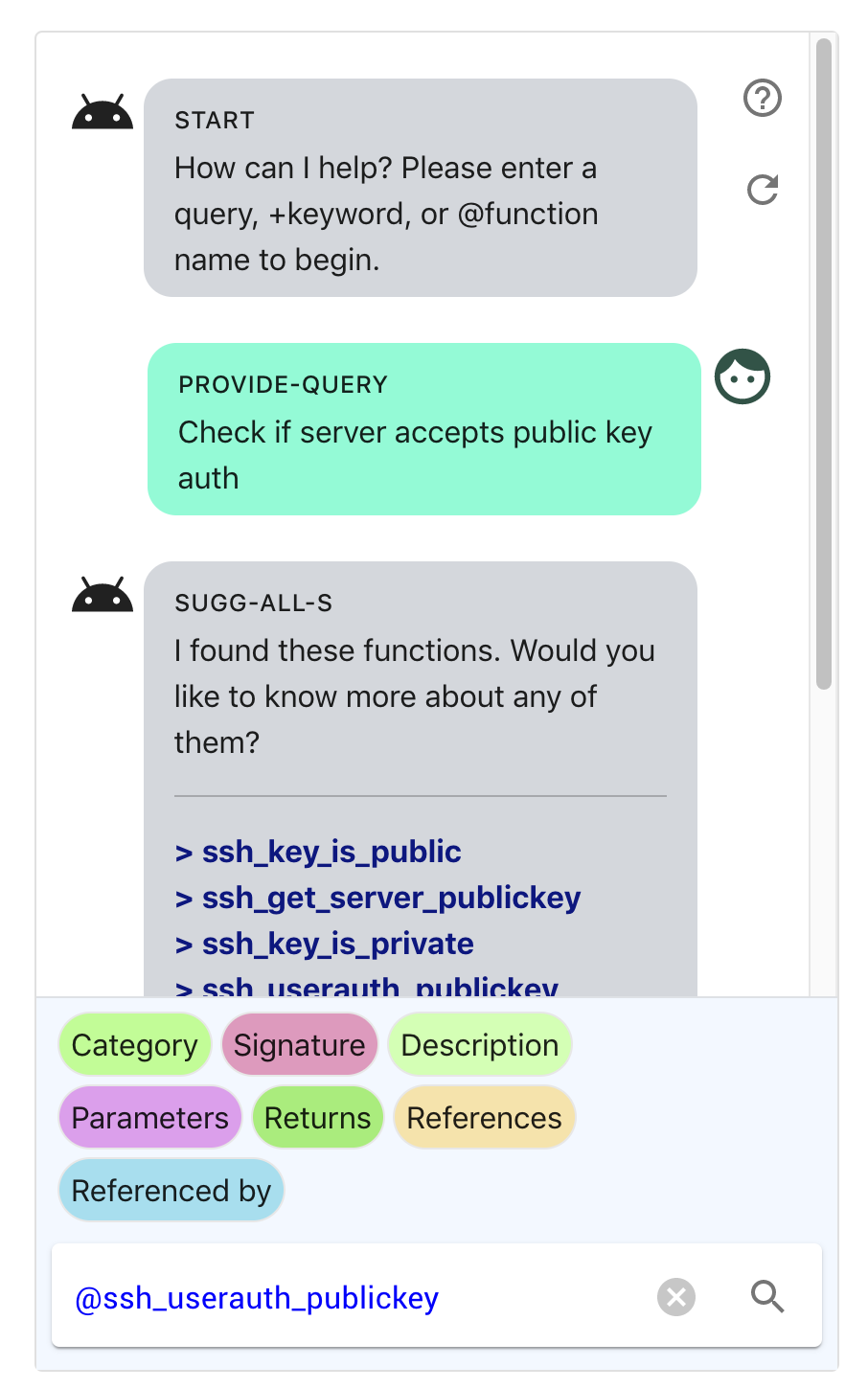}
    \vspace{-0.2cm}
    \caption{The search tool used in our human study. The dialogue system used the same interface for both the learned policy and the single-turn policy.}
    \label{fig:search_tool}
    \vspace{-0.5cm}
\end{figure} 

We created a conversational tool that programmers used to complete their search tasks.  We implemented the tool in Javascript and included it in the study interface. It displayed the same approximate number of messages and amount of text regardless of the size of the user's screen or browser window.

A recommendation by Rieser and Lemon~\cite{rieser2011reinforcement} for dialogue manager evaluation is to minimize the impact of other dialogue system components (e.g. NLU and NLG) by simplifying the input methods and using templated responses for text generation. For user input, we provided a multipurpose search bar. Normal text entered into the search bar was sent to the dialogue manager as a query with the dialogue act type {\small \texttt{provideQuery}}. Users could also add keywords or elicit information about functions by preceding the input with a plus sign (+) or at sign (@), respectively. Additionally, most system messages were accompanied by a set of ``quick response'' bubbles with commands like ``List results'' and ``Next function'' that users could click to invoke specific dialogue acts. For system output, we created a set of response templates covering all system dialogue acts. For example, a message with the {\small \texttt{listResults}} system action would be realized with the text "I found these functions. Would you like to know more about any of them?" followed by a list of $N$ function names, which users could click on to immediately copy to the search bar. 

The tool also included a ``help'' button, which would explain the interface, and a ``restart'' button, which would reset the dialogue state and prompt the user for a new query (the reward function treated a restart as a standard dialogue act).

\subsubsection{API Search Tasks}s
We created six API search tasks using the {\small \texttt{libssh}} API. These tasks asked the programmers to find functions in the API that could be used to implement some high-level functionality e.g., ``Before connecting to an SSH server, specify a host and port for the ssh\_session.'' The complete lists of search tasks are available in our online appendix (see \Cref{sec:reproducibility}).  We based these questions on search tasks used in related literature~\cite{McMillan:2011:PFR:1985793.1985809, hill2014nl,eberhart2020automatically}. Each question targeted a particular function in the API.

\subsubsection{Participants}
We recruited four professional programmers as participants. All were native English speakers with 1-5 years relevant industry experience. Each programmer completed all six search tasks using one of the two policies.

\subsection{Results}

We found that the results during our human study broadly agree with the results from our experiment with synthetic conversations in terms of reward, though provide a different view of success rate.  Consider the following results summary:

\vspace{-0.1cm}
\begin{table}[h!]
\centering
\begin{tabular}{lrr}
Policy                               & learned     & single-turn                            \\ \cline{2-3} 
\multicolumn{1}{l|}{Reward Function} & -12.01 & \multicolumn{1}{r}{-16.69}        \\
\multicolumn{1}{l|}{Success Rate}    & 0.58  & \multicolumn{1}{r}{0.75}           \\ \cline{2-3} 
\end{tabular}
\end{table}
\vspace{-0.1cm}

Two observations stand out.  First, the learned policy outperformed the single-turn strategy in terms of the reward function, with the average reward in general agreement with the previous study (-12.01 for learned in this study versus -13.32 in the previous study, and -16.69 for the single-turn policy versus -15.02).  Second, the success rate for the learned policy is less than single-turn, and is less for both approaches than it was in the study with synthetic conversations.

We attribute these observations in part to the ability of the learned policy to request clarification and recommend individual functions, indicating the system's search confidence to the users and resulting in more concise interactions.  Consider the following summary of the reward function results:

\vspace{-0.15cm}
\begin{table}[h!]
	\centering
	\begin{tabular}{llll}
		Task & learned & single-turn & \textit{difference} \\
		1    & -8.8    & -25.7       & \textbf{16.9}       \\
		2    & -5.8    & -20.0       & \textbf{14.2}       \\
		3    & -15.3   & -9.3        & 6.0                 \\
		4    & -10.3   & -20.2       & \textbf{9.9}        \\
		5    & -19.1   & -13.7       & 5.4                 \\
		6    & -12.8   & -11.4       & 1.4                 \\ \hline
\emph{avg. diff.}   & 13.7	   & 4.3	& 
	\end{tabular}
	
%
	
\end{table}
\vspace{-0.25cm}

The cells in the top table show the average reward function for each task for each policy e.g., -8.8 for the learned policy on task one.  Bold print in the difference column shows tasks for which the learned policy outperformed the single-turn policy.  The average difference is much higher for the tasks when the learned policy is better.  What is happening is that the single-turn policy is just showing a new top-$N$ list each turn, leaving the user to ask a new query each time.  One result is that the learned policy is usually able to find the correct function in fewer turns, and with lower-cost dialogue actions.  However, a side-effect is that the single-turn policy tends to have a higher success rate because it shows a large number of functions (25 turns x 6 functions per turn = 150 potential functions).  This behavior is akin to a brute force search and is likely to achieve success, though at high time cost. 

\subsection{Threats to Validity}

Key threats to validity include 1) participant selection, 2) task design and presentation, and 3) situational factors. Individual differences among programmers and sampling bias may have impacted the observed results. We recruited programmers with similar levels of experience, and had them fill out background surveys to help characterize the participant pool. The specific API search tasks given to programmers may have also impacted the results; it is possible that differently-worded questions or questions targeting different functions could have been more or less difficult for each of the policies. Situational factors, such as time of day, evaluation environment, and distractions may have also impacted individual users' performances. We included the survey and the search tool in the same web interface to help minimize potential distractions that could occur with multiple tabs or windows.

Key threats to external validity include 1) the size and scope of the study, 2) the dialogue management parameters (as discussed in \Cref{sec:ttvquant}), and 3) the formulation of the API search tasks. Ideally, future studies seeking to demonstrate the effectiveness of different policies shoud seek to recruit a larger number of participants and include a broader selection of APIs and search tasks. Furthermore, users interacting with interactive API search systems may not have clearly-defined search tasks as given in this study, and the results here may not generalize to a broader range of API learning tasks.

Key threats to construct validity include 1) the design of the reward function and 2) the design of the dialogue system interface. It is possible that a different reward function would have better captured dialogue manager performance for real users. The interface may have also impacted the users' perceptions of the dialogue manager. While we chose to simplify the system's NLU and NLG to reduce errors irrelevant to the DM, it is possible that implementing the DM in a more-conventional dialogue system would have more accurately mirrored a real-world scenario.

\section{Discussion/Conclusion}
\label{sec:reproducibility}

This paper makes three contributions to the field of software engineering. First, we presented an architecture for dialogue management for API search tasks. We demonstrated how different API search activities can be represented in an action space, as well as how dialogue features and API search results can be represented in a dialogue state. Second, we contributed two dialogue policies for API search that are optimized to minimize the number of dialogue turns required to complete a search task, as well as the amount of information contained in individual messages. Both policies outperformed a baseline policy emulating typical API search engine behavior (see \Cref{sec:quanteval}), suggesting that our policies may provide benefits over naive solutions. Finally, we performed a human study to compare a baseline policy to a dialogue policy derived via reinforcement learning.


For reproducability and future research, we make all source code and experimental materials available online:

{\small \texttt{\url{https://github.com/Zeberhart/dm4api}}}

\bibliographystyle{IEEEtran}
\bibliography{main}

\end{document}